\documentclass[aps,prb,twocolumn, reprint,groupedaddress,amsmath,amssymb]{revtex4-1}

\usepackage{graphicx}
\usepackage{bm}
\usepackage{xcolor}
\begin{document}


\title{Spectral energy analysis of locally resonant nanophononic metamaterials\\ by molecular simulations}


\author{Hossein Honarvar}
\author{Mahmoud I. Hussein}
\email{mih@colorado.edu}
\affiliation{Department of Aerospace Engineering Sciences, University of Colorado Boulder, Boulder, Colorado 80309, USA}


\date{\today}

\begin{abstract}
A nanophononic metamaterial is a new type of nanostructured material that features an array, or a forest, of intrinsically distributed resonating substructures. Each substructure exhibits numerous local resonances, each of which may hybridize with the phonon dispersion of the underlying host material causing significant reductions in the group velocities and consequently a reduction in the lattice thermal conductivity. In this paper, molecular dynamics simulations are utilized to investigate both the dynamics and the thermal transport properties of a nanophononic metamaterial configuration consisting of a freely suspended silicon membrane with an array of silicon nanopillars standing on the surface. The simulations yield results consistent with earlier lattice-dynamics based predictions which showed a reduction in the thermal conductivity due to the presence of the local resonators. Using a spectral energy density approach, in which only simulation data is utilized and no \it{a priori}\rm~information on the nanostructure resonant phonon modes is provided, we show direct evidence of the existence of resonance hybridizations as an inherent mechanism contributing to the slowing down of thermal transport in this system.

\end{abstract}


\maketitle

Material nanostructuring has emerged as a powerful approach in the field of nanoscale heat transfer as it provides a means for direct manipulation of thermal transport properties [\onlinecite{Chen_2000,*cahill2003nanoscale,*Balandin_2005,*cahill2014nanoscale}]. One of the primary applications is thermoelectric materials where there is a need for new concepts and material architectures in order to attain high levels of energy conversion performance [\onlinecite{dresselhaus2007new}]. A promising strategy for increasing the performance of thermoelectric materials, measured by the figure of merit, $ZT$, to levels attractive to industry is to enable a significant lowering of the total thermal conductivity in a manner that does not negatively affect the electrical conductivity and the Seebeck coefficient. In semiconductors, this may be achieved by lowering the lattice thermal conductivity, $k$, by nanostructuring. A common approach is to introduce obstacles, such as holes, inclusions, and interfaces, within a semiconducting material in order to enhance phonon scattering and thus reduce $k$ [\onlinecite{VineisAM2010}]. However, in addition to scattering the phonons, the motion of electrons is likely to be impeded as well by the same obstacles, which diminishes improvements to the $ZT$ value. \\
\indent A new avenue of research for increasing $ZT$ is based on the concept of a \it{nanophononic metamaterial} \rm(NPM) [\onlinecite{PhysRevLett.112.055505}]. In a particular realization of a NPM, presented in Ref.~[\onlinecite{PhysRevLett.112.055505}], an array of nanopillars is built on top of a free-standing membrane or thin film. Each nanopillar exhibits numerous local resonances that roughly span the entire spectrum$-$as many as the number of atoms in the nanopillar multiplied by 3 (the number of degrees of freedom for each atom). This provides millions of resonances for a pillar that is only a few tens of nanometers in size. Each of these resonances may, in principle, couple with the phonon dispersion curves of the underlying membrane and reduce the group velocity significantly at each coupling location in the band structure. This in turn reduces the lattice thermal conductivity, effectively without using scattering as a prime mechanism since there is no longer a need to introduce holes, particles or interfaces into the main body (interior space) of the membrane. Surface scattering is also not required, i.e., the surfaces do not need to be rough. Regardless of the degree of surface roughness, the condition required for the hybridization mechanism to take place is that the phonon mean free path (MFP) distribution be large enough for the resonant standing waves to travel across the full cross-section of the membrane. For a silicon membrane with a thickness on the order of a few nanometers, or a few tens of nanometers, recent experimental studies have shown that the MFP distribution comfortably spans a length scale that at a minimum is on the order of the membrane thickness~[\onlinecite{Johnson_2013,*Minnich_2012,*Neogi_2015}]. Thus resonance hybridization by nanostructuring, as described above, is well suited for thermoelectric energy conversion since it provides a unique opportunity to reduce the lattice thermal conductivity significantly with minimum negative effects on the electrical conductivity and the Seebeck coefficient.

In Ref.~[\onlinecite{PhysRevLett.112.055505}], the thermal conductivity of the pillared silicon membrane was obtained using a model based on the Boltzmann transport equation (BTE) following the single-mode relaxation time (SMRT) approximation~[\onlinecite{ashcroft_1976,*Srivastava_1990}]. In this model, complete phonon dispersion information was provided from harmonic lattice dynamics (LD) calculations of the NPM unit cell incorporating the motion of all atoms and using an empirical interatomic potential. The scattering time constants were obtained by a fitting procedure on available experimental thermal conductivity data for suspended uniform silicon membranes. The analysis was conducted for room temperature conditions. The results for the particular unit-cell configuration examined showed that the nanopillars reduce the thermal conductivity of an otherwise uniform membrane by a factor of 2. In this paper, we investigate a very similar pillared silicon membrane configuration using equilibrium molecular dynamics (MD) simulations and spectral energy density calculations. Our results confirm the validity of the earlier lattice-dynamics based predictions, and, importantly, provide direct evidence of the resonance-hybridization phenomenon taking place within the MD simulations.

\begin{figure}[t] 
	\includegraphics{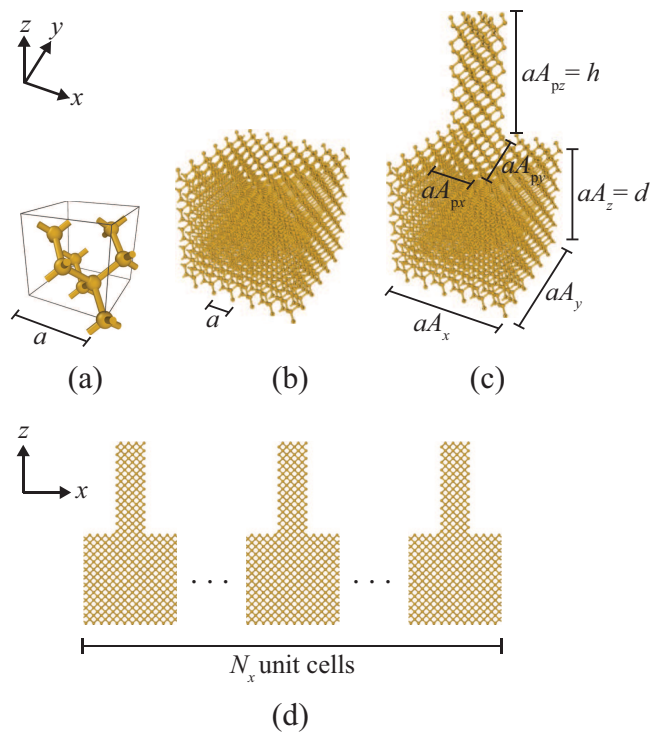}
	\caption{\label{fig:fig_1} 
		(a) Silicon conventional cell with the lattice constant $a=5.431~\textnormal{\AA}$, (b) uniform (unpillared) membrane unit cell (c) NPM (pillared) unit cell, (d) array of NPM unit cells forming an MD computational domain.}
\end{figure}
\begin{figure}[t]
 \includegraphics{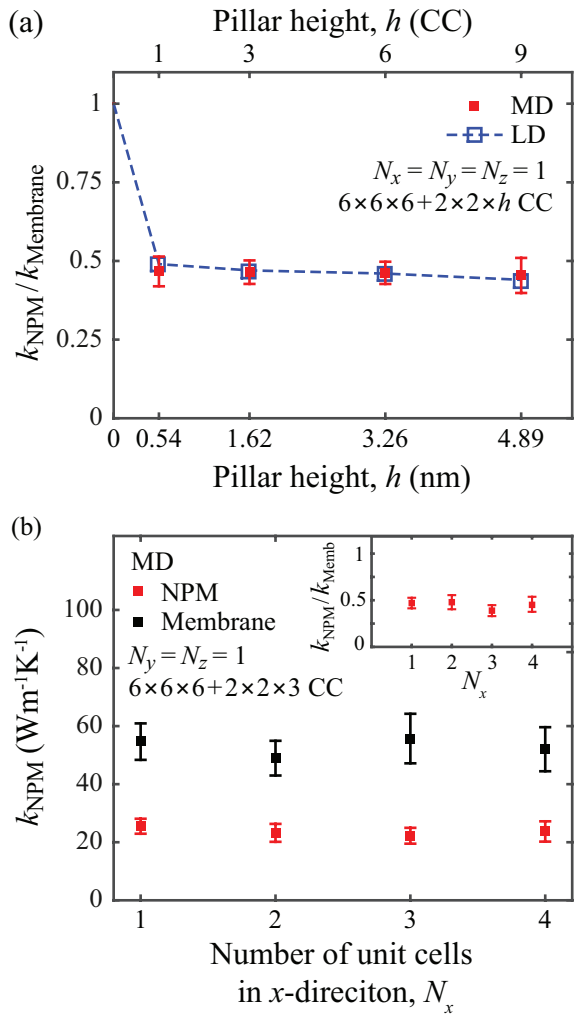} 
\caption{\label{fig:fig_2} Thermal conductivity predictions at $T=300$ K. In (a) and in the inset of (b), the NPM thermal conductivity, $k_{\text{NPM}}$ is normalized with respect to the thermal conductivity of a uniform membrane with the same thickness, $k_{\text{Membrane}}$. In (a), $N_{x} = N_{y} = N_{z} = 1$ and the unit cell has dimensions $ 6\times 6\times 6+2\times 2\times h$ CC. In (b) $N_{y} = N_{z} = 1$, $N_{x}$ is varying and the unit cell has dimensions $ 6\times 6\times 6+2\times 2\times 3$ CC. The error bars represent uncertainties in the MD-based results.}
\end{figure}

We consider a unit-cell model of a suspended pillared membrane consisting of $A_{x}\times A_{y}\times A_{z}$ conventional cells (CC) of silicon forming the base (membrane portion) and $A_{\text{p}x}\times A_{\text{p}y}\times A_{\text{p}z}$ CC of silicon forming the nanopillar. A silicon CC consists of eight atoms packaged as a cube with a side length of $a=0.5431$ nm (see Fig.~\ref{fig:fig_1}a). Thus, the unit cell considered has a membrane thickness of $d=aA_{z}$ and a nanopillar height of $h=aA_{\textbf{p}z}$. An LD calculation is by construction performed on a single unit cell. However, in an MD simulation, it is a matter of choice on how many unit cells to include in the model to provide an adequate representation of the physical phenomenon of interest. In general, we describe an array of unit cells by $N_{x}\times N_{y}\times N_{z}$, where $N_{i}$ denotes the number of unit cells in the $i$th-direction.\\
\indent Figure~\ref{fig:fig_1}c displays a particular configuration we consider featuring a unit cell with dimensions of $aA_{x}=aA_{y}=aA_{z}=6$ CC $=3.258$~nm, $aA_{\text{p}x}=aA_{\text{p}y}=2$ CC $=1.086$~nm and $aA_{\text{p}z}=6$ CC $=3.258$~nm. A shorthand notation is adopted in the rest of the paper such that this unit cell's dimensions are represented as `$6\times6\times6+2\times2\times 6$' CC. The model for a corresponding uniform thin membrane (i.e., with the nanopillar removed) is shown in Fig.~\ref{fig:fig_1}b. For the LD calculations, Bloch conditions are applied at the membrane in-plane boundaries for the pillared and the uniform unit cells, whereas the bottom and top membrane surfaces and all the nanopillar surfaces are free. For the MD simulations, where the computational domain consists of one or more unit cells, standard periodic boundary conditions are applied at the in-plane boundaries. For all models and calculations, room temperature, $T = 300$ K, is assumed and the Stillinger-Weber empirical potential is used to represent the interatomic forces~[\onlinecite{stillinger1985computer}]. Only wave propagation and phonon transport along the $x$-direction ($\Gamma X$ path in reciprocal lattice space) is considered. \\ 
\indent In addition to the models shown in Figs.~\ref{fig:fig_1}b and 1c, we consider cases where the nanopillar height is $A_{\text{p}z}=1, 3$ and 9 CC. Using the same procedure and fitting parameters as in Ref.~[\onlinecite{PhysRevLett.112.055505}] (i.e., prediction of the thermal conductivity using BTE with SMRT approximation, the phonon dispersion by harmonic LD calculations~[\onlinecite{gale2003general}], and the lifetimes using experimentally fitted scattering constants based on uniform membranes with the same thickness), we obtain the results shown in blue (dashed line) in Fig.~\ref{fig:fig_2}a. We observe that (i) the nanopillars reduce $k$ by nearly a factor of 2, as in Ref.~[\onlinecite{PhysRevLett.112.055505}], and (ii) the increasing height of the nanopillars causes a modest, but increasing, reduction in $k$. Next we execute a series of equilibrium MD simulations on the same cases and analyze the results using the Green-Kubo (GK) formulation to predict the thermal conductivities~[\onlinecite{Zwanzig_1965,Ladd_1986,*volz2000molecular,*Che_2000,*Schelling_2002,*mcgaughey2004thermal,Landry2008}]. The advantage of MD simulations is that they are inherently anharmonic thus the phonon lifetimes are directly and implicitly incorporated. Classical MD simulations like the ones we conduct here are valid at room temperature where quantum effects are negligible. The GK approach follows the linear response theorem and is based on the dissipation of equilibrium fluctuations for the heat current vector, $\bm{J}$. Defining the time average of the heat current auto-correlation function (HCACF) for a generally anisotropic material as $\langle{\bm{J}(0)}\otimes{\bm{J}(t)}\rangle$, where $\otimes$ denotes the tensor product, the thermal conductivity tensor is expressed as
\begin{equation}
{\bm{k}} = \frac{1}{k_{B} VT^{2}}\int{\langle{\bm{J}(0)}\otimes{\bm{J}(t)}\rangle \textrm{d} t,}
\end{equation}
where $k_{B}$ is the Boltzmann constant, $V$ is the system volume and $T$ is the temperature.

The MD simulations are performed using the LAMMPS software [\onlinecite{plimpton1995fast}]. The systems are equilibrated under $NVE$ (constant mass, volume and energy) ensembles with a time step ${\Delta}{t}=0.5$ fs for a time span of 6 ns, which is sufficiently large compared to the longest phonon lifetime. The HCACFs converge within the first 500 ps. All reported thermal conductivities are the average of values from six independent simulations with different initial velocities. Furthermore, the thermal conductivities are averaged in the $x$- and $y$-directions effectively resulting in an averaging over twelve predicted values.

The MD-based thermal conductivities are also plotted in Fig.~\ref{fig:fig_2}a and are in excellent agreement with the LD-based predictions, thus supporting the conclusions presented in Ref.~[\onlinecite{PhysRevLett.112.055505}]. We point out again to the slight decrease in the thermal conductivity with increasing nanopillar height, which provides a promising avenue for future design studies to achieve further reductions in the thermal conductivity.

In the results of Fig.~\ref{fig:fig_2}a, the MD computational domain consists of only a single unit cell with standard periodic boundary conditions applied, i.e., $N_{x} = N_{y} = N_{z} = 1$. This assumption is made in light of our interest in the effects of local resonances on an intrinsic in-plane material property, the lattice thermal conductivity. A larger computational domain, however, is needed in order to account for Bragg scattering stemming from the periodic presence of the nanopillars$-$which is a wave interference mechanism on the order of the unit cell size. In Fig.~\ref{fig:fig_2}b, we show thermal conductivity predictions for a unit cell with dimensions $6\times6\times6+2\times2\times 3$ CC within a computational domain that consists of $N_{y} = N_{z} = 1$ and an array of $N_{x}$ cells laid out periodically in the $x$-direction. In principle, one might expect an additional reduction in the thermal conductivity due to Bragg scattering; however, in our system, this additional reduction$-$if it exists$-$appears to be very modest that it falls within the margin of error of the processed data. Furthermore, the results indicate that there is no noticeable computational size effect when simulating a domain consisting of only a single unit cell with standard boundary conditions applied; this in fact is one of the benefits of equilibrium MD simulations [\onlinecite{Landry2008}]. 

In addition to altering the dispersion curves and creating resonance hybridizations that roughly span the full phonon spectrum, the nanopillars are expected to also trigger additional phonon-boundary scattering compared to a uniform membrane. In order to examine whether the resonance hybridization mechanism does indeed exist and unfolds within the MD simulations (and as suggested by the harmonic lattice dynamics band diagrams), we compute the spectral energy density (SED)~[\onlinecite{thomas2010predicting, mcgaughey2014predicting, wang2014atomistic, larkin2014comparison}], which is a quantity obtained directly from the simulations.\\
\indent There are two SED formulations reported in the literature for phonon transport problems. In one SED expression, referred to as ${\Phi}$, the MD atom velocities are projected onto the phonon normal modes of the constituent unit cell, which are obtained separately from lattice dynamics calculations. This approach allows for an accurate prediction of both phonon frequencies and lifetimes~[\onlinecite{larkin2014comparison}]. In an alternative formulation, the SED expression requires knowledge of only the crystal unit-cell structure and does not require any \it a priori \rm knowledge of the phonon mode eigenvectors. This alternative SED expression, referred to as ${\Phi'}$, accurately predicts only the phonon frequencies and not the lifetimes~[\onlinecite{larkin2014comparison}]. In the current investigation, we intentionally seek a technique that allows us to extract the frequency-wave vector spectrum directly from the MD simulations without any prior knowledge of the phonon band structure. Thus the ${\Phi'}$ formulation is perfectly suited for our aim.   

As provided in Ref.~[\onlinecite{thomas2010predicting}], the SED expression, ${\Phi'}$, is a function of wave vector, $\bm{\kappa}$, and frequency, $\omega$, and for our all-silicon system is given by 

\begin{equation}
{{{\Phi'}}\left( \bm{\kappa}, \omega \right)} = \mu_{0} 
\displaystyle\sum\limits_{\alpha}^{3}
\displaystyle\sum\limits_{b}^{n}
\begin{vmatrix}
{\displaystyle\sum\limits_{l}^{N}{ \displaystyle\int\limits_{0}^{\tau_{0}}{\dot{{u}}_{\alpha}
\left (\scriptsize{\!\!\!\!
\begin{array}{l l} 
\begin{array}{l} l\\b
\end{array} \!\!\!\!\!\!
 &;~ t
\!\!\!\!\end{array}}
\right )
e^{{\textrm{i}}\left[\bm{\kappa}.\bm{r}_{0}
\left (\!\!
\scriptsize{\begin{array}{l} l\\0
\end{array} \!\!}
\right )-wt\right]} \textrm{d} t}} }
\end{vmatrix}
^{2} 
\label{eq:SED}
\end{equation}
where $\mu_{0}=m/(4\pi \tau_{0} N)$, $m$ is the mass of a silicon atom, $\tau_{0}$ is the total simulation time, ${\bm{r}}_{0}$ is the equilibrium position vector of the $l$th unit cell, and ${\dot{{u}}}_{\alpha}$ is the $\alpha$-component of the velocity of the $b$th atom in the $l$th unit cell at time $t$. There are a total of $N=N_{x}\times N_{y}\times N_{z}$ unit cells in the simulated computational domain with $n$ atoms per unit cell. 

\begin{figure*}[t]
	\includegraphics{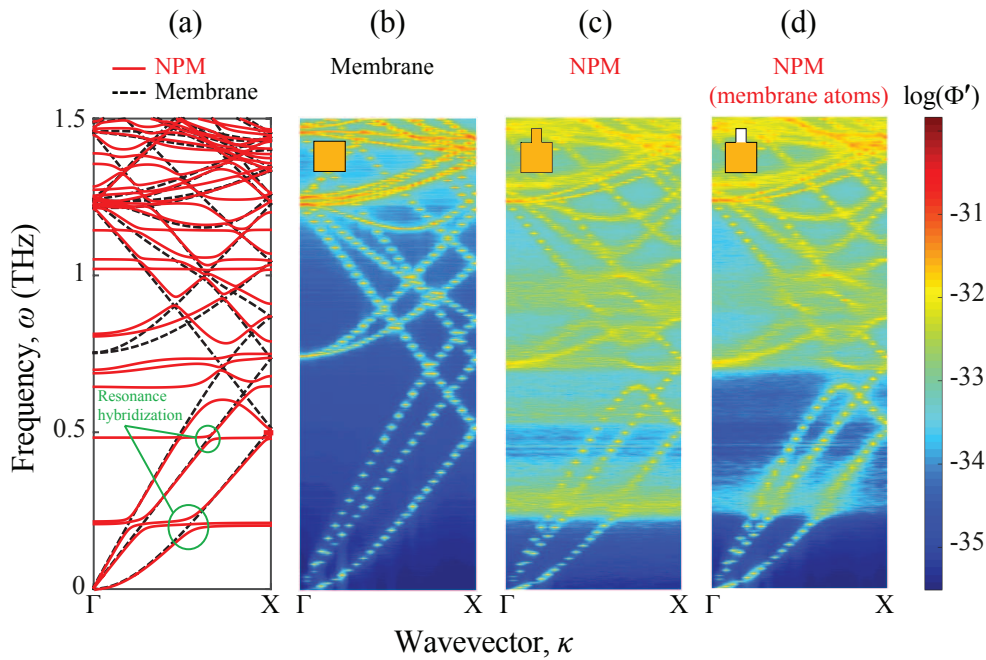}
	\caption{\label{fig:fig_3} Phonon dispersion of a NPM and a corresponding uniform membrane with the same thickness. Subfigure (a) shows the phonon dispersion for both material structures as directly obtained from harmonic LD calculations. Subfigure (b) and (c) show the SED spectrum for the uniform membrane and NPM, respectively. The SED spectrum of the NPM considering only the membrane atoms is displayed in (d). The NPM unit-cell  dimensions are $6\times 6\times 6+2\times 2\times 3$ CC; the uniform membrane unit-cell dimensions are $6\times 6\times 6$ CC. Each inset presents a schematic of unit cell analyzed.}
\end{figure*}

We note that in Eq.~(\ref{eq:SED}) the phonon frequencies can only be obtained at the set of allowed wave vectors as dictated by the crystal structure. For our model, the $\Gamma X$-path wave vectors are $\kappa_{x}={2\pi j}/({N_{x}A_{x}})$, $j=0$ to $N_{x}/2$. We consider a unit cell identical to the one depicted in Fig.~\ref{fig:fig_1}c except for the nanopillar height which we select to be $h=3$ CC. For the computational domain, we set $N_{x} = 50$ and $N_{y}= N_{z}=1$, which gives a $\Gamma X$ wave-vector resolution of ${\Delta}{\kappa}_{x}=0.02$. MD simulations under $NVE$ conditions are executed for this system for $2^{22}$ time steps and based on ${\Delta}{t}=0.5$ fs as earlier. Equation~(\ref{eq:SED}) is evaluated by computing the Fourier transform of the velocity trajectories extracted every $2^{5}$ steps. 

The results from these calculations are remarkable. As a reference, the phonon band structure as obtained from standard harmonic LD calculations is shown in Fig.~\ref{fig:fig_3}a for the NPM and, for comparison, for a uniform membrane with the same $d$. The corresponding SED spectrum is shown in Fig.~\ref{fig:fig_3}b for the uniform membrane and in Fig.~\ref{fig:fig_3}c for the NPM. Only the frequency range $0 \leq \omega \leq 1.5$ THz is shown because higher frequencies are difficult to distinguish in the SED field. The phonon dispersion emerging from the MD simulations matches very well with that obtained by the independent LD calculations, thus providing confidence in both sets of simulations/calculations.~\footnote{The close resemblance of the LD and SED phonon dispersion curves indicates that the degree of anharmonicities in the MD simulations is relatively modest at the temperature considered} In particular, the first set of nanopillar local resonances, present at nearly 0.2 THz, are clearly observed in the NPM SED spectrum, appearing as horizontal lines. More importantly, the interaction of these resonances with the acoustic branches of the underlying membrane are distinctly observed and follow exactly the hybridization profiles featured in the LD dispersion curves. Resonance hybridizations are also clearly observed at higher frequencies where local resonance modes interact with optical dispersion branches. \\
\indent Given our interest in the effects of the nanopillar resonances on the heat carrying phonons within the base membrane, we recalculate the SED spectrum for the NPM considering only the contributions of the atoms housed in the base membrane (i.e., discounting the SED contributions of the nanopillar atoms). The outcome of this calculation is shown in Fig.~\ref{fig:fig_3}d where we see direct evidence that the nanopillar resonances alter the fundamental nature of the phonon traveling waves within the membrane. These alterations result in a significant reduction in the phonon group velocities at each location in the band structure where an interaction takes place, which in turn leads to a reduction in the lattice thermal conductivity as seen from the GK analysis presented in Fig.~\ref{fig:fig_2}. While phonon-phonon and phonon-boundary scatterings are still important mechanisms in the membrane-based systems we have considered, the results are consistent with the understanding that the MFP distribution comfortably spans, at a minimum, the length scale of the membrane thickness. Such a MFP distribution is sufficiently broad to allow at least a portion of the nanopillar standing waves to impact the group velocities of in-plane traveling phonons across the entire cross-section of the membrane.       

This research benefited from the support of the NSF CAREER grant number 1254931. The authors wish to acknowledge Dr. Bruce Davis and Dr. Lina Yang for fruitful discussions concerning the thermal conductivity prediction methods used in this study. This work utilized the Janus supercomputer, which is supported by the NSF (award number CNS-0821794) and the University of Colorado Boulder (CU-Boulder). The Janus supercomputer is a joint effort of CU-Boulder, the University of Colorado Denver and the National Center for Atmospheric Research. 
\bibliography{SED_Ref}

\end{document}